\documentclass[journal, twoside, web]{IEEEtran}

\usepackage{cite}
\usepackage{dblfloatfix} 
\usepackage{algorithmic}
\usepackage{graphicx}
\usepackage{textcomp}
\def\BibTeX{{\rm B\kern-.05em{\sc i\kern-.025em b}\kern-.08em
    T\kern-.1667em\lower.7ex\hbox{E}\kern-.125emX}}
\usepackage[utf8]{inputenc}
\usepackage[english]{babel}
\usepackage{amsmath, amsfonts, amssymb}
\usepackage{soul}

\usepackage{array}
\usepackage{multirow}

\newcolumntype{C}[1]{>{\centering\arraybackslash}p{#1}}
\newcolumntype{L}[1]{p{#1}}

\DeclareFontFamily{U}{BOONDOX-calo}{\skewchar\font=45 }
\DeclareFontShape{U}{BOONDOX-calo}{m}{n}{
  <-> s*[1.05] BOONDOX-r-calo}{}
\DeclareFontShape{U}{BOONDOX-calo}{b}{n}{
  <-> s*[1.05] BOONDOX-b-calo}{}
\DeclareMathAlphabet{\mathcalboondox}{U}{BOONDOX-calo}{m}{n}
\SetMathAlphabet{\mathcalboondox}{bold}{U}{BOONDOX-calo}{b}{n}
\DeclareMathAlphabet{\mathbcalboondox}{U}{BOONDOX-calo}{b}{n}

\DeclareFontFamily{OT1}{pzc}{}
\DeclareFontShape{OT1}{pzc}{m}{it}{<-> s * [1.000] pzcmi7t}{}
\DeclareMathAlphabet{\mathpzc}{OT1}{pzc}{m}{it}

\usepackage{multidef}
\usepackage{mathtools}

\multidef[prefix=set]{\ensuremath{\mathbb{#1}}}{A-Z}

\multidef[prefix=cal]{\ensuremath{\mathcal{#1}}}{A-Z}

\multidef[prefix=v]{\ensuremath{\mathbf{#1}}}{a-z}

\multidef[prefix=m]{\ensuremath{\mathbf{#1}}}{A-Z}

\multidef[prefix=layer]{\ensuremath{\mathpzc{#1}}}{a-z}

\newcommand{\trace}{\mathrm{tr}}

\newcommand{\erdosrenyi}{Erdős–Rényi }

\usepackage{booktabs}
\usepackage{tabularx}
\newcolumntype{Y}{>{\centering\arraybackslash}X}

\begin{document}

\title{Community Detection in Multi-frequency EEG Networks}
\author{Abdullah Karaaslanli, Meiby Ortiz-Bouza, Tamanna T. K. Munia, and Selin Aviyente
\thanks{This work was in part supported by the National Science Foundation grant CCF 2006800.}
\thanks{The authors are with the Department of Electrical and Computer Engineering, Michigan State University, East Lansing, MI 48824 USA (e-mail: karaasl1@msu.edu; ortizbou@msu.edu; aviyente@egr.msu.edu).}}

\maketitle

\begin{abstract}
\textit{Objective:} In recent years, the functional connectivity of the human brain has been studied with graph theoretical tools. One such approach is community detection which is fundamental for uncovering the localized networks. Existing methods focus on networks constructed from a single frequency band while ignoring multi-frequency nature of functional connectivity. Therefore, there is a need to study multi-frequency functional connectivity to be able to capture the full view of neuronal connectivity. \textit{Methods:} In this paper, we use  multilayer networks to model multi-frequency functional connectivity. In the proposed model, each layer corresponds to a different frequency band. We then extend the definition of modularity to multilayer networks to develop a new community detection algorithm. \textit{Results:} The proposed approach is applied to electroencephalogram data collected during a study of error monitoring in the human brain. The differences between the community structures  within and across different frequency bands for two response types, i.e. error and correct, are studied. \textit{Conclusion:}  The results indicate that following an error response, the brain organizes itself to form communities across frequencies, in particular between theta and gamma bands while a similar cross-frequency community formation is not observed for the correct response. Moreover, the community structures detected for the error response were more consistent across subjects compared to the community structures for correct response. \textit{Significance: } The multi-frequency functional connectivity network models combined with multilayer community detection algorithms can reveal changes in cross-frequency functional connectivity network formation across different tasks and response types.
\end{abstract}


\section{Introduction}
\label{sec:introduction}

The human brain consists of various units which interact with each other through structural and functional links. Advances in functional and structural neuroimaging technology allow the brain to be modeled as a network using graph theoretic tools. In this modeling, the nodes correspond to the different brain units and the edges represent structural or functional connections among these units \cite{bullmore2009complex}. In order to characterize the topology and dynamics of brain networks, various descriptive and inferential network measures are utilized; such as centrality, degree distribution and small-worldness \cite{boccaletti2006complex, muldoon2016small, mattar2016flexible, bassett2015learning, bassett2017network} with respect to disease, task, learning, cognitive control, attention and memory \cite{bullmore2009complex, braun2015dynamic, bassett2017network, mattar2016flexible, cole2007cognitive, berger2000pathologies, tessitore2012default, seeley2009neurodegenerative}. 

Current network models have been mostly limited to examining a single network instance either of a subject, a frequency band or a task. However, most neurophysiological recordings, such as the electroencephalogram (EEG), allows one to capture brain dynamics across multiple temporal and spatial scales. Reducing this rich information into a single network disregards the high amount of dependency that exists between networks of different subjects, frequency bands and time points. Thus, a principled mathematical framework to accurately study this multiplicity of brain connectivity is needed. 

Recently, multilayer networks \cite{kivela2014multilayer, de2013mathematical, boccaletti2014structure} have been proposed as a mathematical framework to study multiple networks simultaneously. Multilayer networks consist of multiple layers, each of which carry information from a different network while the interactions between layers represent the dependency between these networks. Due to their ability to represent and study multi-dimensional and multi-scale data, multilayer networks gained attention in network neuroscience \cite{de2017multilayer, muldoon2016network, vaiana2018multilayer}. To this end, measures that are developed to characterize the layered structure of multilayer networks are employed to study brain connectivity, allowing for a richer analysis than prior single network based studies \cite{de2016mapping}. 

\begin{figure*}[t]
    \centering
    \includegraphics{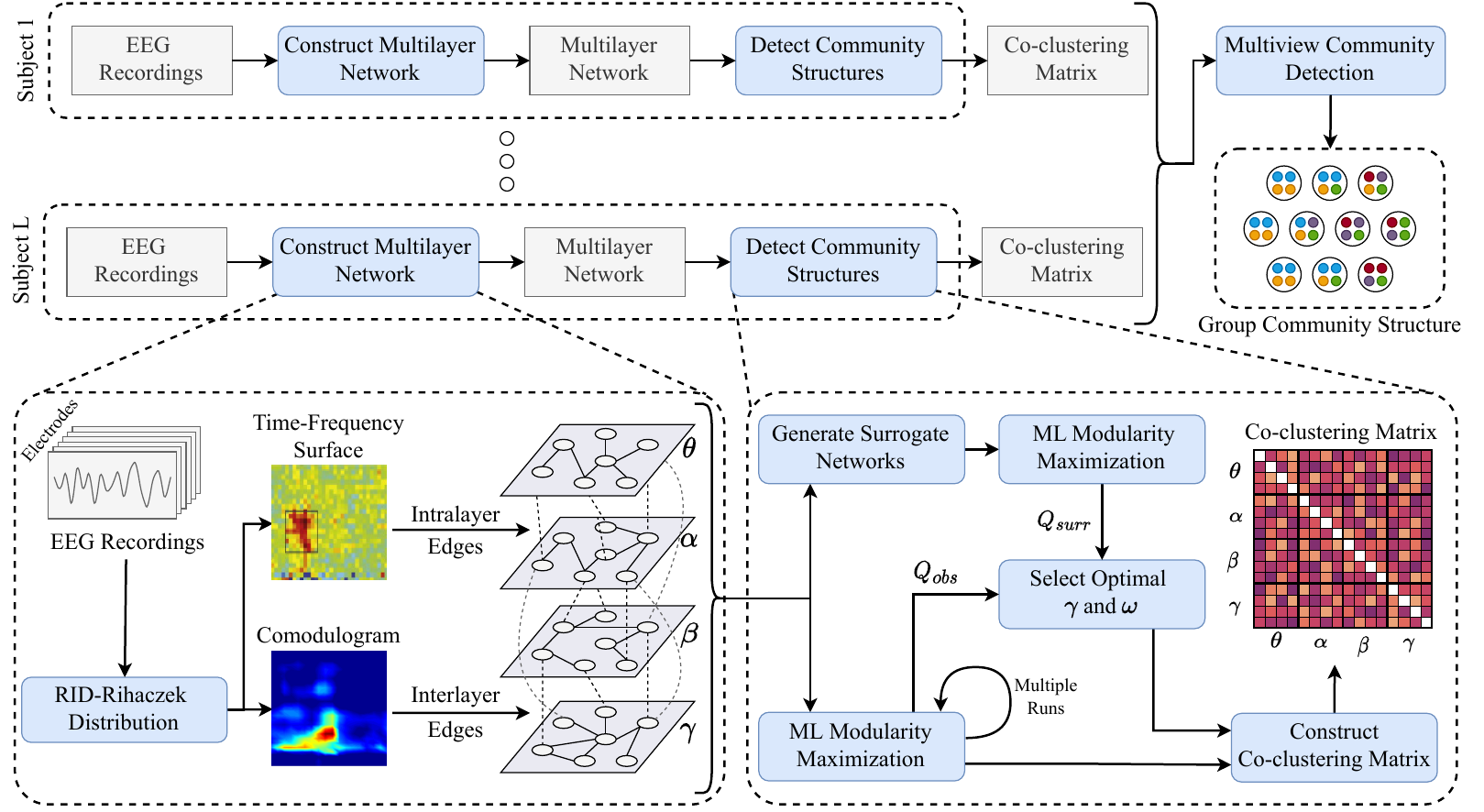}
    \caption{Flowchart of the proposed approach for community detection of 
    multi-frequency EEG networks. Bottom two panels illustrate multilayer 
    network construction (left) and community detection for each subject (right).}
    \label{fig:flowchart}
\end{figure*}

In this paper, we aim to characterize the topological organization of multilayer brain networks by developing a multilayer community detection algorithm. The proposed approach is outlined in Fig. \ref{fig:flowchart}. The main contributions of the proposed framework are: 
\begin{itemize}
\item First, we introduce a signal processing based approach for constructing multi-frequency networks from EEG data. In this approach, the intra- and inter-layer edges are quantified by novel time-frequency based phase synchrony and phase amplitude coupling (PAC) measures, respectively. Thus, the constructed network is a \textit{general multilayer network} with inter-layer edges allowed between all brain regions.
\item Second, modularity function is defined for multilayer networks based on a multilayer null model that preserves the layer-wise node strengths while randomizing the remaining characteristics of the network. The corresponding modularity maximization algorithm is capable of detecting communities that are both common across a subset of layers and private to each layer.
\item Third, once the multilayer community structure for each subject is detected, a consensus clustering approach based on multiview modeling is used to obtain group community structure of a set of subjects. In the multiview network model, each layer corresponds to a subject's co-clustering matrix constructed from community structures obtained from  multiple runs of modularity maximization. Multilayer spectral clustering is then applied to the resulting multiview network. This approach considers all of the subjects simultaneously and detects a common community structure across subjects summarizing their network topology.
\item Finally, this is the first work of its kind that considers a general multilayer network perspective for EEG data and uses the community structure to characterize the brain activity across multiple frequencies, simultaneously. In particular, the group level differences between the two response types during Flanker task, i.e., error and correct, are evaluated from a multi-frequency network perspective. 
\end{itemize}

\section{Related Work}
\label{sec:related_work}

\subsection{Multilayer Brain Networks}

Recently, the study of brain networks has undergone a process of adapting classical single-layer functional connectivity network concepts to a more general multilayer description \cite{de2017multilayer, betzel2017multi, vaiana2018multilayer, de2016mapping, tewarie2016integrating}. The meaning of layer can vary depending on context; including multiple modalities, subjects, time points, and frequency bands. Some examples include multi-modal networks where each layer represents a different modality. For example, \cite{battiston2017multilayer} study a two layer network that combines structural and functional networks of a subject, which are constructed from the subject's diffusion tensor imaging (DTI) data and funtional MRI (fMRI), respectively. Two layers are connected with inter-layer edges, which link a brain region in the functional network to itself in the structural network. Another example is the temporal network where each layer represents connectivity over some time period \cite{bassett2015learning, braun2015dynamic}. The inter-layer edges are usually allowed only between consecutive time periods and between nodes that represent the same brain regions. Temporal brain networks are employed to analyze how the community structure and the network topology of functional networks change over time.

More recently, functional multilayer networks are constructed, where layers correspond to well-known frequency bands at which the brain operates \cite{brookes2016multi, tewarie2016integrating, dang2020rhythm, buldu2018frequency, yu2017selective}. Compared to single-layer networks, which do not distinguish the contributions coming from different frequency bands and consider a single frequency band, multilayer networks allow the study of connectivity across multiple frequency bands simultaneously. For instance, recent work has considered multi-frequency functional connectivity networks for resting state fMRI \cite{gohel2015functional, zou2019multi, sasai2014frequency, sasai2021frequency}. These studies have shown that the network organization such as hubs may differ across different frequency bands \cite{sasai2014frequency}. More recently, multi-frequency networks were considered in neurophysiological studies \cite{dimitriadis2021assessing, tewarie2016integrating, brookes2016multi, puxeddu2021comprehensive}. Brookes et al. \cite{brookes2016multi} use magnetoencephalographic (MEG) recordings to construct functional multilayer networks, where each layer corresponds to the links within a given frequency band, and the inter-layer edges correspond to the cross-frequency coupling across frequency bands resulting in a general multilayer network. It is shown that there is statistically significant difference between supra-adjacency matrices of the multilayer networks of  control and schizophrenia subjects. Similar multi-frequency networks are constructed from resting state MEG recordings of subjects affected by Alzheimer's disease. Analyzes of these networks shows the abnormal distribution of regional connectivity across frequency bands in unhealthy subjects compared to healthy individuals, revealing an abnormal loss of inter-frequency centrality in memory-related association areas \cite{guillon2017loss}. More recently, a multi-frequency network model was considered for EEG data \cite{puxeddu2021comprehensive, puxeddu2019optimal}, where a multilayer network is constructed by concatenating the functional connectivity matrices for 40 different frequency bins. Generalized Louvain algorithm for multiview networks is then employed to detect the community structure of these multi-frequency networks. However, in this work, the inter-layer connections were only allowed between nodes corresponding to the same brain areas, implying that the network is not a general multilayer network.

\subsection{Community Detection in Multilayer Networks}
Community detection is one of the main network analysis tools and has been applied to several complex
systems, including biological and social \cite{fortunato2016community}. In single-layer networks, communities are defined as groups of nodes that are more
strongly connected among themselves than they are to the rest of the nodes, while in multilayer networks,
these groups of nodes are shared across multiple layers. Traditional community detection methods usually need modifications to achieve a good performance on a multilayer network.

Community detection methods for multilayer and multiview networks can be categorized into the following classes \cite{magnani2021community}; optimization of quality functions, e.g., multilayer modularity index \cite{pramanik2017discovering, zhang2017modularity} and normalized cut \cite{chen2017block,deford2019spectral,chen2017block2}, extensions of stochastic block model (SBM) to multiview and multilayer networks \cite{han2015consistent,barbillon2017stochastic,stanley2016clustering}, methods based on random walks \cite{de2015identifying}, label propagation  \cite{boutemine2017mining}, and nonnegative matrix factorization \cite{al2022community,ma2018community}.  

In \cite{pramanik2017discovering}, the definition of modularity is extended for 
multilayer networks. The proposed modularity index is then maximized using   
Girvan-Newman and Louvain algorithms. In \cite{chen2017block,chen2017block2}, the 
normalized cut measure is extended to multiview and multilayer networks by 
constructing a block Laplacian matrix where each block corresponds to a layer. Then they apply spectral clustering to this matrix for finding the communities.  In \cite{tiomoko2018latent}, a Weighted Stochastic 
Block Model (WSBM) is proposed to detect communities on each layer and also the ones that are shared across layers of a heterogeneous weighted network . However, this method only detects common communities if they are shared by all of the layers. In \cite{de2015identifying}, 
a method based on random walks, Multiplex-Infomap, is proposed. Although this method 
identifies intra- and inter-layer communities in multilayer networks, it 
usually groups physical nodes across layers in the same community. In 
\cite{al2022community}, the authors propose an approach based on nonnegative matrix 
factorization that detects intra- and inter-layer communities in general multilayer 
networks. However, this method is only applicable to networks with two layers. In 
\cite{boutemine2017mining}, a community detection method based on Label Propagation Algorithm (LPA) is 
proposed for multilayer networks. LPA follows the intuition 
that a label would get trapped inside a group of densely connected nodes and end up 
having the same label indicating they belong to the same community. This method simultaneously
identifies the communities and the subset of layers where these communities belong to. Although these methods are developed networks with multiple layers, only a few of them are applicable to general multilayer networks 
where connections may exist between different physical nodes  across layers 
\cite{pramanik2017discovering,chen2017block2,al2022community}.

\section{Background}
\label{sec:background}

\subsection{Multilayer Networks}
\label{ssec:multilayer_networks}

An undirected single-layer network $G=(V, E, \mA)$ is represented by a node 
set $V$ with $|V|=N$, an edge set $E \subseteq V\times V$, where $e_{uv}$ are 
associated with weights $w_{uv}$ and adjacency matrix $\mA \in \setR^{N\times N}$
with $A_{uv} = A_{vu} = w_{uv}$ if $e_{uv} \in E$ and $0$, otherwise. The strength
of node $u$ is the total weight of the edges incident to it, i.e. $s_{u} = 
\sum_{v=1}^n A_{uv}$.

An undirected multilayer network is a quadruplet $\calM = (\calV, \calL, V, E)$
where $\calV$ is the set of physical entities, $\calL$ is the set of layers with
$|\calL| = L$ \cite{kivela2014multilayer}. $V \subseteq \calV \times \calL$ with
$|V|=N$ is the set of nodes, which are representations of physical entities in
layers and $E \subseteq V\times V$ is the edge set. In this paper, nodes of a 
multilayer network are represented as $u^\layerh$, where $u \in \calV$ and 
$\layerh \in \calL$. An edge between $u^\layerh$ and $v^\layerk$ is denoted by 
$e_{uv}^{\layerh\layerk}$ and associated with the weight $w_{uv}^{\layerh\layerk}$. 
$V$ can be partitioned into layers, that is $V=\bigcup_{\layerh=1}^L V^\layerh$ 
where $V^\layerh$ is the set of nodes in layer $\layerh$ with $|V^{\layerh}| = 
N^{\layerh}$. Similarly, $E$ can be partitioned as $E = \bigcup_{\layerh=1}^L 
E^{\layerh}\ \cup \ \bigcup_{\layerh \neq \layerk=1}^L E^{\layerh\layerk}$, 
where $E^\layerh$ is the intra-layer edges of layer $\layerh$ and $E^{\layerh\layerk}$ 
is the inter-layer edges between nodes in layers $\layerh$ and $\layerk$. Using 
this notation, one can define intra-layer graphs $G^{\layerh}=(V^{\layerh}, 
E^\layerh)$ and bipartite inter-layer graphs $G^{\layerh\layerk} = (V^{\layerh}, 
V^{\layerk}, E^{\layerh\layerk})$. $\calM$ is represented by a supra-adjacency 
matrix $\mA \in \setR^{N\times N}$, defined as:
\begin{align}
    \mA = \begin{bmatrix}
              \mA^{1} & \mA^{12} & \dots & \mA^{1L} \\
              \mA^{21} & \mA^{2} & \dots & \mA^{2L} \\
              \vdots & \vdots & \ddots & \vdots \\
              \mA^{L1} & \mA^{L2} & \dots & \mA^{L}
          \end{bmatrix},
\end{align}
where $\mA^{\layerh}$ is the adjacency matrix of $G^\layerh$ and 
$\mA^{\layerh\layerk}$ is the incidence matrix of the bipartite graph, 
$G^{\layerh\layerk}$. Layer-wise strength of a node $u^\layerh$ is the sum 
of weights corresponding to the edges connected to the nodes in layer $\layerk$, 
i.e., $s_{u^\layerh}^\layerk = \sum_{v\in V^{\layerk}} w_{uv}^{\layerh\layerk}$ 
with $w_{uv}^{\layerh\layerk} = 0$ if $e_{uv}^{\layerh\layerk} \not\in
E^{\layerh\layerk}$ ($E^{\layerh}$ if $\layerh = \layerk$).

Multilayer graphs can be categorized based on the structure of their inter-layer 
graphs. In this paper, we consider two types of multilayer graphs. The first one is
multiview graphs, where inter-layer edges are only allowed between nodes 
representing the same physical entities. The second type is general multilayer 
graphs, which do not have any restrictions on the structure of 
$G^{\layerh\layerk}$'s. In the following, general multilayer graphs will be 
referred to as multilayer graphs for simplicity. 

\subsection{Modularity}
\label{ssec:modularity}

Given an undirected single-layer graph $G=(V, E, \mA)$, community detection aims 
to partition the node set $V$ into $K$ communities, $\{C_1, \cdots, C_K\}$. This 
is usually achieved by optimizing a quality function that quantifies the goodness 
of a given partition \cite{fortunato2010community, fortunato2016community}. Among 
various objective functions, the most commonly  used one is modularity 
\cite{newman2004finding}, which quantifies the quality  of a partition by 
comparing the intra-community edge density to that expected under a null model and
is calculated as follows:
\begin{align}
    \label{eq:modularity}
    Q = \sum_{i=1}^N \sum_{j=1}^N (A_{ij} - P_{ij}) \delta_{g_ig_j},
\end{align}
where $P_{ij}$ is the expected edge weight between nodes $i$ and $j$ under the 
null model, $g_i$ is the community of node $i$, and $\delta_{g_{i}g_{j}} = 1$ if 
$g_i=g_j$ and $0$, otherwise. Different null models can be used to define $P_{ij}$ 
depending on the graph under study. The most commonly used null models are the 
configuration and \erdosrenyi null models \cite{reichardt2006statistical}. 

Modularity as defined in \eqref{eq:modularity} suffers from the resolution limit 
\cite{fortunato2007resolution}, i.e., it cannot detect small communities. 
Therefore, it has been extended to include a resolution parameter $\gamma$ 
\cite{reichardt2006statistical}:
\begin{align}
    \label{eq:modularity_with_resolution_parameter}
    Q = \sum_{i=1}^N \sum_{j=1}^N (A_{ij} - \gamma P_{ij}) \delta_{g_ig_j}.
\end{align}
By tuning $\gamma$, one changes the resolution of the modularity function such 
that larger $\gamma$ values can detect smaller communities. 

\section{Multilayer EEG Networks}
\label{sec:multilayer_eeg_networks}

\subsection{Data Acquisition}
\label{ssec:data_acquisition}

The EEG data was acquired during a cognitive control-related error processing task where the subjects performed a letter version of the speeded reaction Flanker task \cite{moran2012sex}. The experimental protocol of this study was approved by the Institutional Review Board (IRB) of the Michigan State University (IRB: LEGACY13-144). The data collection was conducted by following the regulations approved by this protocol. Prior to data acquisition, all subjects signed an informed and written consent form. During recording, each subject was presented with a string of five letters at each trial. Letters could be congruent (e.g., SSSSS) or incongruent stimuli (e.g., SSTSS) and the subject was instructed to respond to the center letter with a standard mouse. The trials started with a flanking stimulus (e.g., SS SS) of 35ms followed by the target stimuli (e.g., SSSSS/SSTSS) displayed for about 100 ms. The total display time is 135 ms, followed by a 1200 to 1700 ms inter-trial break between the trials. These trials capture the Error-Related Negativity (ERN) after an error response and the Correct-Related Negativity (CRN) after a correct response. As earlier studies suggested a rise in synchronization related to ERN for the 25–75 ms time window \cite{ozdemir2015hierarchical}, all the analysis was conducted for the 25–75 ms time period of the data. For each subject, 480 total trials (each of 1-second in duration) were recorded, where the number of error trials varied from 20 to 61 across the subjects. For a fair comparison between ERN/CRN, the same number of correct trials were selected randomly. The EEG signals were recorded with a BioSemi ActiveTwo system using a cap with 64 Ag–AgCl electrodes placed at standard locations of the International 10–20 system. The sampling rate of the data was 512 Hz. After cleaning the artifacts, volume conduction was minimized using the Current Source Density (CSD) Toolbox \cite{tenke2012generator}. A multilayer network with four layers is constructed for each subject and each response type (error vs. correct) where layers correspond to the four EEG frequency bands: $\theta$ (4-7 Hz), $\alpha$ (8-12 Hz), $\beta$ (13-30 Hz), $\gamma$ (31-100 Hz). In this paper, we consider data from 20 participants.

\subsection{Intra-layer Edges}
\label{ssec:intralayer_edges}

The intra-layer edges are computed using a reduced interference Rihaczek (RID-Rihaczek) time-frequency distribution-based phase synchrony index known as RID Rihaczek time-frequency phase synchrony (RID-TFPS) \cite{aviyente2011phase, aviyente2011time}. Compared to existing measures like wavelet transform, RID-TFPS provides uniformly high time and frequency resolution and has been reported as a robust phase synchrony index for noisy signals  \cite{aviyente2011time}. For an arbitrary signal $x(t)$, the RID-Rihaczek distribution is obtained as \cite{aviyente2011time}:
\begin{align}
\begin{split}
\label{eq:rid_rihaczek}
    C(t,f) = \iiint
        & e^{-(\theta\tau)^2/\sigma} e^{j\theta\tau/2}     
            x(u+\tau/2)x^{*}(u-\tau/2) \\ 
        & e^{-j(\theta{t}+2\pi{f\tau}-\theta u)}{du}{d\tau}{d\theta},
\end{split}
\end{align}
where $e^{-(\theta\tau)^2/\sigma}$ refers to the Choi-Williams kernel and $e^{j\theta\tau/2}$ defines the kernel function for the Rihaczek distribution \cite{rihaczek1968signal}. $C(t,f)$ belongs to Cohen’s class of distribution and thus satisfies the marginals and preserves the energy. This complex time-frequency distribution is utilized to calculate the phase difference $\phi_{u,v}(t,f)$, between two signals $x_u$ and $x_v$ as:
\begin{align}
\label{eq:phasediff}
    \phi_{u,v}(t,f) = \arg \bigg[\frac{C_u(t,f) C_{v}^*(t,f)}{| C_u(t,f) | | C_v(t,f) |}\bigg].
\end{align}

The phase synchrony between two brain locations measures the consistency of the phase differences $\phi_{u,v}(t,f)$ across trials and is quantified using the Phase Locking Value (PLV) index. PLV measures the degree of phase locking by computing inter-trial variability of the phase differences as follows  \cite{lachaux1999measuring}:
\begin{align}
\label{eq:plv}
     \text{PLV}_{u,v}(t,f) = \frac{1}{K}{\bigg| \sum_{k=1}^{K} e^{j\phi_{u,v}^k(t,f)}\bigg | },
\end{align}
where $K$ is the total number of trials and $\phi_{u,v}^{k}(t,f)$ is the phase 
difference between $x_u^{k}$ and $x_v^{k}$ as given in 
\eqref{eq:phasediff} for trial $k$. After computing the pairwise PLV values between all possible brain locations, the average pairwise synchrony within a predefined time window of interest, 
$W = [t_{1},t_{2}]$, and a chosen frequency band is used as intra-layer edge weights, 
i.e., $w_{uv}^{\layerh\layerh}=\frac{1}{|W|}\frac{1}{|\layerh|}\sum_{t \in W}\sum_{f 
\in \layerh} \text{PLV}_{u,v} (t,f)$, $1\leq u,v \leq N$, where $N$ is the number of brain locations and  and $|\layerh|$ is the bandwidth of the particular frequency band $\layerh$.

\subsection{Inter-layer Edges}
\label{ssec:interlayer_edges}

The inter-layer edges are calculated using a form of cross-frequency coupling known as PAC, which computes the modulation of the amplitude/power of a high frequency rhythm along the phase of a slower frequency rhythm \cite{bragin1995gamma,tort2010measuring}. In this study, PAC between two brain locations is estimated by applying a RID-Rihaczek time-frequency-based PAC approach \cite{munia2019time, munia2019comparison}. To quantify PAC, first we obtain analytic signals $C_u(t,f)$ and $C_v(t,f)$ at node $u$ and $v$, respectively, by using RID-Rihaczek distribution as defined in \eqref{eq:rid_rihaczek}. These analytic signals are utilized to extract the instantaneous high frequency amplitude envelope $a^{u}_{f_a}(t)$ at node $u$, and the instantaneous low frequency phase $\phi_{f_p}^{v}(t)$ at node $v$, where $f_{p}$ and $f_{a}$ are respective frequencies within the $\layerh$th and $\layerk$th frequency bands. For node $u$, $a^{u}_{f_a}(t)$ is obtained from the frequency constrained time marginal of $C_u(t,f)$ as:
\begin{align}
\label{eq:amp_envlp}
    a^{u}_{f_a}(t) = \int\nolimits_{f_{a_1}}^{f_{a_2}} {C_{u}(t,f)df},
\end{align}
where $f_{a_1}$  and $f_{a_2}$ is the bandwidth around the chosen high frequency. Similarly, the low frequency phase components at node $v$ is obtained from $C_v(t,f)$, as:
\begin{align}
\label{eq:pac_phase}
    \phi^{v}_{f_p}(t) = \arg \left[\frac{C_{v}(t,{f_p})}{| C_{v}(t,{f_p}) |}\right]. 
\end{align}

Once the amplitude and phase components are extracted, PAC is estimated by distributing $a^{u}_{f_a}(t)$ and $\phi^{v}_{f_p}(t)$ to a composite vector in the complex plane at each time point and measuring the amplitude normalized modulation index (MI) proposed in \cite{ozkurt2011critical} as follows:
\begin{align}
\label{eq:MI}
    \text{MI}_{{u,v}}(f_{p},f_{a},t) = \frac{1}{\sqrt{K}} \frac{\bigg| 
         {\sum_{k=1}^{K}{a^{u,k}_{f_a}(t)}{e^{j\phi^{v,k}_{f_p}(t)}}}
    \bigg|}{\sqrt{ {\sum_{k=1}^{K}{a^{u,k}_{f_a}(t)}^2}}}.
\end{align}
As the amplitude of the signal normalizes the computed MI, the MI value is scaled between 0 and 1 \cite{hulsemann2019quantification}. Thus, the inter-layer edges between node $u$ and $v$ are constructed as $w_{uv}^{\layerh\layerk} = 
\frac{1}{|W|}\frac{1}{|\layerh||\layerk|}\sum_{t \in W}\sum_{f_{p}\in
\layerh}\sum_{f_{a}\in \layerk}MI_{u,v}(f_{p},f_{a},t)$.

\section{Multilayer Community Detection}
\label{sec:multilayer_community_detection}

\subsection{Multilayer Modularity}
\label{ssec:multilayer_modularity}

\begin{figure*}[t]
    \centering
    \includegraphics[width=\textwidth]{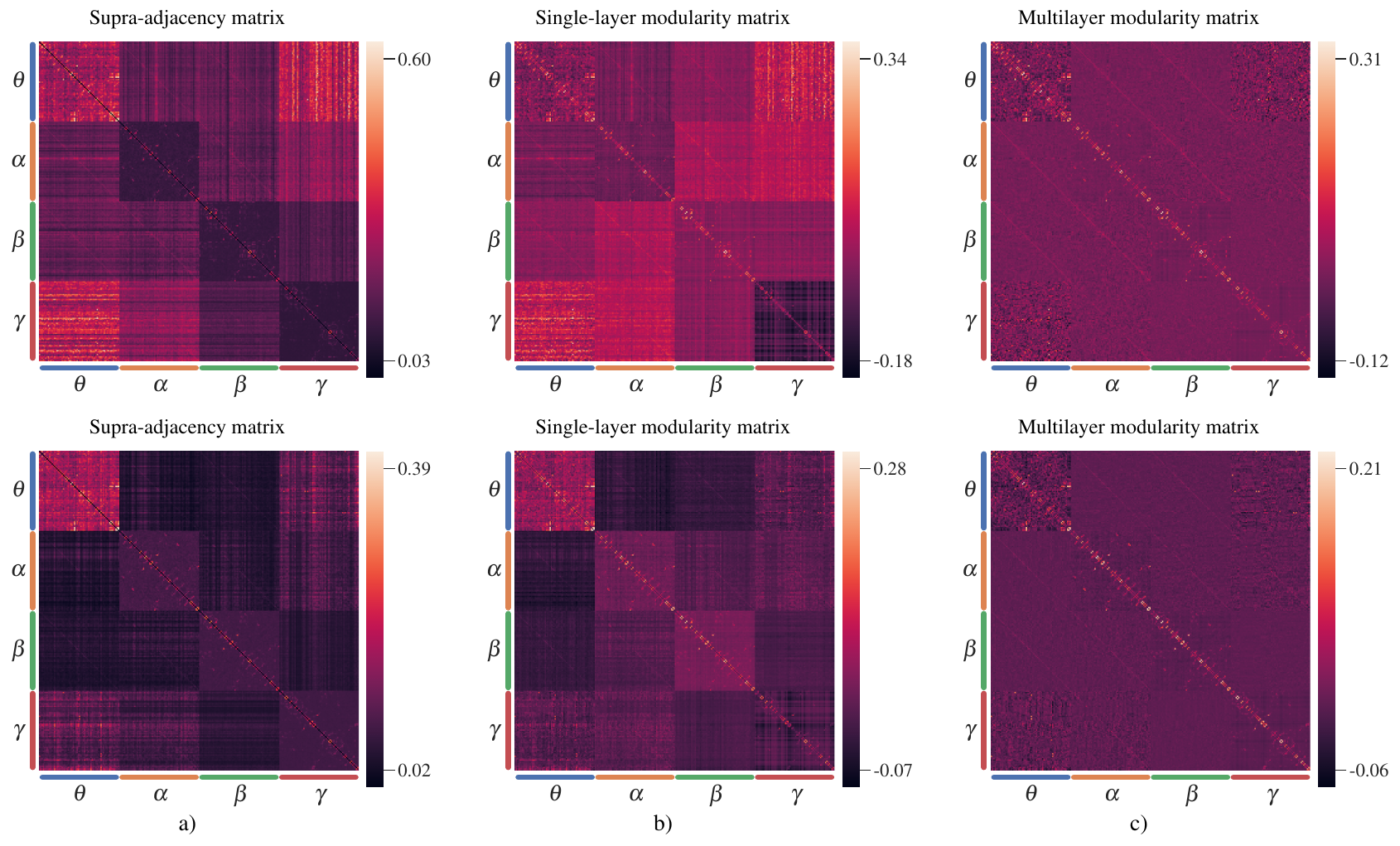}
    \caption{Supra-adjacency and two types of modularity matrices for multilayer EEG 
    networks for a single subject. Top row corresponds to the error response, while 
    the bottom row corresponds to the correct response. a) Supra-adjacency matrices, 
    b)  Modularity matrices obtained when the multilayer network is considered as a 
    single-layer network, c) Modularity matrices obtained using the modularity 
    function defined in \eqref{eq:multilayer_modularity}. Rows and columns of the 
    matrices are color coded to indicate the layers corresponding to the nodes.}
    \label{fig:mlgraph_matrices}
\end{figure*}

Modularity function compares intra-community edges of the observed network to those 
of a null model to quantify the quality of a community structure. The null model is 
a random graph model with some properties, \textit{e.g.} edge density, of the 
observed network. The properties to be preserved are determined such that they 
are assumed to be unrelated to the community structure \cite{fornito2016fundamentals}.
For example in the  configuration null model, the degree of each node is preserved 
so that the identified community structure would not be affected by the heterogeneity
of the degree distribution. This assumption is based on the fact that nodes with 
a high degree tend to connect with each other merely because they have high 
number of connections and not necessarily because they are within the same 
community \cite{karrer2011stochastic}. To prevent this tendency to bias 
community detection, the null model preserves the node degrees. On the other 
hand, \erdosrenyi null model does not make such an assumption and allows the
identified community structure to be influenced by the degree distribution.

Based on this insight on the role of null models, we extend the 
definition of modularity function to multilayer networks by considering which
properties of the observed multilayer network we want to preserve in the 
null model. In Fig. \ref{fig:mlgraph_matrices}a, the supra-adjacency matrices 
of the multilayer EEG networks are shown for one subject for both error and 
correct responses. From this figure, it can be seen that the edge weights are
heterogeneous across layers for both response types. For instance, the 
intra-layer edges for $\theta$ band and inter-layer edges between $\theta$ and 
$\gamma$ bands are very strong, while intra-layer connections for $\alpha$, 
$\beta$ and $\gamma$ bands are weaker for error response. Any community detection 
method applied to this network without taking this heterogeneity into account 
would likely partition the nodes based on the layer label rather than the true 
community membership. In order to prevent this, the null model used in the 
definition of the modularity function should preserve the heterogeneity of edge 
weights across layers as follows. 

\paragraph*{Multilayer Configuration Null Model} Let $\calM$ be a multilayer 
network with $m^{\layerh\layerh}$ defined as the total weight of the intra-layer 
edges in layer $\layerh$ and $m^{\layerh\layerk}$ defined as the total weight of the 
inter-layer edges between layers $\layerh$ and $\layerk$. The multilayer configuration 
null model preserves layer-wise node strengths while randomizing the remaining 
characteristics of $\calM$. The expected edge weight between $u^\layerh$ and 
$v^\layerk$ is then:
\begin{align}
    \label{eq:multilayer_configuration_null_model}
    P_{uv}^{\layerh\layerk} = \frac{s_{u^\layerh}^\layerk s_{v^\layerk}^\layerh}{(1+\delta_{\layerh\layerk}) m^{\layerh\layerk}},
\end{align}
where $\delta_{\layerh\layerk} = 1$ if $\layerh=\layerk$ and $0$, otherwise.

The multilayer modularity is then defined as follows:
\begin{align}
    \label{eq:multilayer_modularity}
    Q = \sum_{\layerh=1}^L \sum_{\layerk=1}^L \omega^{\layerh\layerk} \sum_{i=1}^{n^\layerh} \sum_{j=1}^{n^\layerk} (A_{ij}^{\layerh\layerk} - \gamma^{\layerh\layerk} P_{ij}^{\layerh\layerk}) \delta_{g_i^\layerh g_j^\layerk},
\end{align}
where $\gamma^{\layerh\layerk}$'s are the resolution parameters corresponding to  intra- and
inter-layer graphs and $\omega^{\layerh\layerk}$'s are the scaling parameters that 
weigh the importance of different intra- and inter-layer connections. 
In the following, we use a common resolution parameter across layers, i.e. 
$\gamma^{\layerh\layerk} = \gamma$. Scaling parameters are set as
$\omega^{\layerh\layerk} = \omega$ if $\layerh \neq \layerk$ and $1$, otherwise. 
\eqref{eq:multilayer_modularity} can be optimized with greedy algorithms, 
such as the Louvain algorithm \cite{blondel2008fast}, developed for maximizing the
single-layer modularity function defined in \eqref{eq:modularity}. In this work, 
we used the Leiden algorithm, which is an extension of the Louvain algorithm with better 
performance \cite{traag2019louvain}.

Since the proposed multilayer configuration null model preserves the heterogeneity of 
edge weights across layers, the corresponding modularity matrix should 
not have the heterogeneity observed for the original supra-adjacency matrices in Fig. 
\ref{fig:mlgraph_matrices}a.  Fig. 
\ref{fig:mlgraph_matrices}c. shows the modularity matrices, where the
heterogeneity across layers is reduced. On the other hand, Fig. 
\ref{fig:mlgraph_matrices}b. shows the modularity matrices computed with the single-layer configuration 
null model. In this case, the heterogeneity across layers is very similar to that of the
supra-adjacency matrix. These results indicate that multilayer configuration null 
model defined in \eqref{eq:multilayer_configuration_null_model} preserves the
heterogeneity of edge weights across layers as desired.

While recent work \cite{pramanik2017discovering} has considered a similar definition of 
modularity for multilayer networks, in that work, the resolution parameter is set to 
1 and the inter-layer scale parameters are determined from  $m^{\layerh\layerk}$.
However, the resolution limit of modularity function necessitates one to consider
different resolution parameters to obtain better community structures \cite{sporns2016modular}. Moreover, as our 
results show, considering different values for inter-layer scale can reveal 
important aspects of multilayer community structure. Therefore, in this work, 
we do not fix these parameters and try to learn their optimal values as described in Results section.

\subsection{Group Community Structure}
\label{ssec:group_community_structure}

\begin{figure*}[t]
    \centering
    \includegraphics{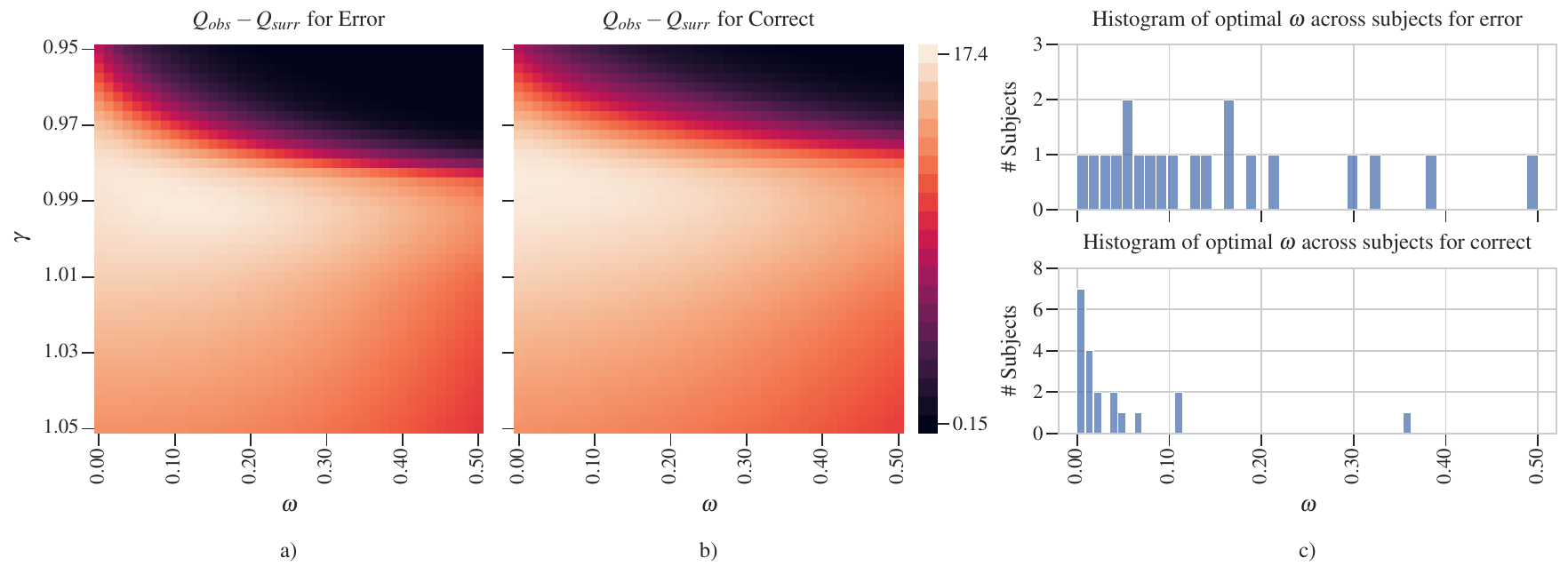}
    \caption{Selection of the resolution ($\gamma$) and inter-layer scale ($\omega$) parameters: a) and b) show the average of $Q_{obs} - Q_{surr}$ across 20 subjects for error and correct responses, respectively. c) shows the histogram of optimal $\omega$ values  for error (top) and correct (bottom) responses across subjects.}
    \label{fig:parameters}
\end{figure*}

Once the community structures of the multilayer networks for a group of subjects
are detected, it is often desirable to find a group community structure, which
summarizes the shared communities across subjects. There are two widely used 
approaches in network neuroscience to find group community structure 
\cite{betzel2017multi}. The first approach is based on selecting a representative 
subject and using this subject's community structure as the group structure 
\cite{doron2012dynamic}. Representative subject is selected as the one whose 
community structure has the greatest average similarity to the other subjects' 
community structures. Since this approach ignores the community structure of 
all but one subject, it generally does not yield satisfactory results. The second 
approach is consensus clustering \cite{lancichinetti2012consensus}, which
combines information from multiple community structures to derive the group 
community structure. This approach constructs a co-clustering matrix $\mA$ 
where $A_{uv}$ is the number of times nodes $u$ and $v$ are in the same community
across all subjects and applies community detection to $\mA$ to find the group 
structure. However, this method is highly dependent on the quality of the  detected 
community structure for each subject. Since  modularity maximization is an 
NP-hard problem \cite{brandes2007modularity}, modularity maximization algorithms 
yield locally optimal results. Furthermore, modularity function has a strong 
degeneracy \cite{good2010performance}, i.e., multiple community structures with 
similar modularity values may be significantly different from each other. 
Therefore, there can be a diverse set of informative community structures for each 
subject and relying on a single community structure from each subject  can be 
problematic for consensus clustering \cite{kirkley2022representative}. 

In order to address these issues, in this paper we propose a group community
structure detection method based on multiview graphs. Given $L$ subjects, for 
each subject we maximize the modularity function with the optimal $\gamma$ and 
$\omega$ values 100 times to obtain 100 community structures. From these community 
structures, we construct a co-clustering matrix $\mA^\layerh$ for each subject 
$\layerh \in \{1,2,\ldots,L\}$ as described above. This process results in $L$ 
co-clustering matrices, which can be modeled as the layers of a multiview graph, 
where each layer is an undirected, weighted graph corresponding to each subject. 
The group community structure can be found from this multiview graph.

Multiview community detection methods generally fall into two categories 
\cite{magnani2021community}. The first category of methods find a single 
common community structure using information across all layers. The second 
category of methods finds a community structure for each 
layer while regularizing these structures to have some similarity. Since our 
goal is to find the group community structure of $L$ subjects, we focus on the 
first category. In particular, we use Spectral Clustering on Multi-Layer graphs 
(SC-ML) \cite{dong2013clustering}, which finds the multiview community structure
by applying spectral clustering to a modified Laplacian defined as:
\begin{align}
    \mL_{mod} = \sum_{{\layerh}=1}^L \mL^\layerh 
        - \alpha \sum_{\layerh=1}^L \mU^\layerh {\mU^{\layerh}}^\top ,
\end{align}
where $\mL^{\layerh}$ is the the normalized graph Laplacian defined as  $\mL^{\layerh}=(\mD^{\layerh})^{-1/2}(\mD^{\layerh}-\mA^{\layerh})(\mD^{\layerh})^{-1/2}$,
 where $\mD^{\layerh}$ is the diagonal matrix of node strengths and $\mU^{\layerh}$ is
the low-rank subspace embedding of layer $\layerh$. In this work, 
we set $\alpha=0.5$, following the guidelines in \cite{dong2013clustering}. 

\section{Results}
\label{sec:results}

\subsection{Resolution Parameter and Inter-layer Scale Selection}

We propose a statistical testing approach comparing the modularity value of the 
observed multilayer network to that of surrogate networks to determine the resolution 
and inter-layer scale parameters in \eqref{eq:multilayer_modularity}. Such methods 
have been previously used in network neuroscience literature for selecting the 
resolution parameter in single-layer networks and coupling parameter in multiview 
graphs \cite{bassett2013robust, sporns2016modular, betzel2017multi}. Given an 
observed multilayer network $\calM$ and $c$ surrogate multilayer networks, we perform 
community detection on surrogate multilayer networks for a given pair of $\gamma$ and 
$\omega$ values, and calculate the modularity values of the detected community 
structures. Let $Q_{surr}$ be the average of these $c$ modularity values. Next, we 
perform modularity maximization for $\calM$  $c$ times and compute the average of the 
modularity values for the $c$ community structures, $Q_{obs}$. This process is 
repeated for different pairs of $\gamma$ and $\omega$ values, and the pair with 
the largest difference, $Q_{obs} - Q_{surr}$, is selected as the optimal parameter 
values. 

The surrogate networks are generally constructed from the observed graph by 
randomly swapping the edges while preserving some properties, \textit{e.g.} node 
strength, of the observed graph \cite{maslov2002specificity, ansmann2011constrained}. 
Since the multilayer EEG networks  are fully connected and 
weighted, we focus on randomization techniques presented in
\cite{ansmann2011constrained}, where two approaches are proposed for fully 
connected single-layer networks. The first technique swaps edges while preserving 
the edge weights. The second approach randomly modifies the edge weights 
while preserving node strengths. Both edge weights 
and node strengths cannot be preserved at the same time when randomizing fully 
connected weighted networks, as this "in most cases only leaves one possible 
surrogate network, namely, the original network" \cite{ansmann2011constrained}. 
We employed the first technique, as the second one may result in surrogate 
networks with unrealistic edge weights for EEG networks, \textit{e.g.} edge weights 
larger than 1. Thus, surrogate multilayer networks are generated as follows: we 
select two edges $e_{uv}^{\layerh\layerk}$ and $e_{st}^{\layerl\layerm}$ and swap 
their edge weights. Edges are selected such that $\layerh = 
\layerl$ and $\layerk = \layerm$, which ensures that the heterogeneity of edge
weights across layers is preserved.

We first study the optimal values of $\gamma$ and $\omega$ obtained by the above
procedure. For each subject and each response type, 100 surrogate networks are 
generated and their community structures are found for each $(\gamma, \omega) \in 
\Gamma \times \Omega$, where $\Gamma = \{\gamma : \gamma = 0.95 + 0.0025n, n \in \{0, 
1, \dots, 40\}\}$ and $\Omega = \{\omega : \omega = 0.0 + 0.0125n, n \in \{0, 1, 
\dots, 40\}\}$. For each subject, 100 community structures are detected for each 
$(\gamma, \omega) \in \Gamma \times \Omega$. Modularity values of these community 
structures are evaluated and the optimal $\gamma$ and $\omega$ values for each subject
are then found from $Q_{obs} - Q_{surr}$.  

Figs. \ref{fig:parameters}a. and  \ref{fig:parameters}b. show the average of 
$Q_{obs} - Q_{surr}$ across subjects for error and correct responses, respectively. 
For both response types, optimal $\gamma$ is found to be close to $0.99$, while 
optimal $\omega$ values are observed to be more diverse across subjects, ranging 
between 0.0-0.2 for error and between 0.0-0.1 for correct. In Fig. 
\ref{fig:parameters}c, we plotted the histogram of the optimal $\omega$ values 
across subjects. This figure shows that the optimal $\omega$ values are non-zero 
for all subjects except one for the error response. On the other hand, for correct 
response, the optimal $\omega$ values are close to $0$ for 7 subjects, while most 
of the remaining subjects have optimal $\omega$ values close to $0$. This result 
indicates that inter-layer connections are not important for the community structure 
of correct response, as $\omega=0$ means inter-layer part of the modularity function 
is equal to zero. For error response, inter-layer edges are influential in community 
formation indicating the importance of cross-frequency coupling. This result is 
in line with prior work showing increased cross-frequency coupling between high-frequency oscillations related to motor activity and to visual processing in the gamma band and low frequency cognitive control signals which are activated after an error response \cite{munia2019time}. It has been suggested that low frequency network oscillations in prefrontal cortex, e.g. $\theta$ band, guide the expression of motor-related activity in action planning \cite{riddle2021causal}.

\subsection{Consistency of Community Structures for Error and Correct}

\begin{figure}[t]
    \centering
    \includegraphics{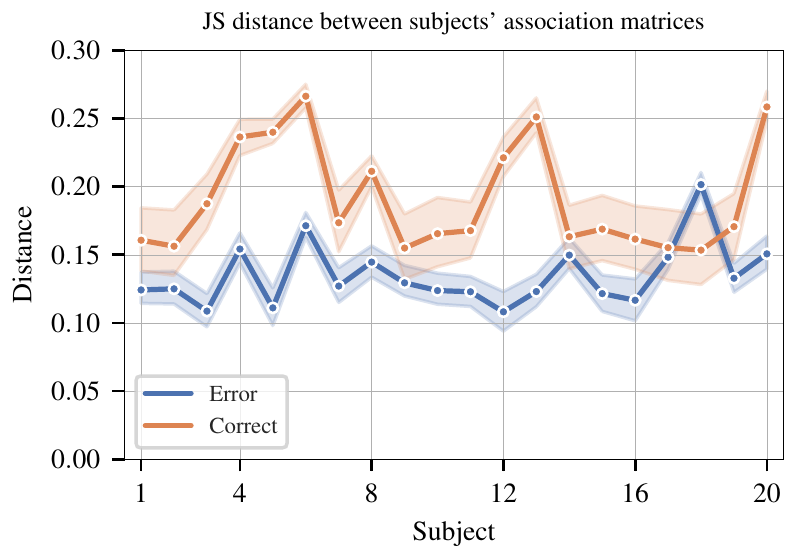}
    \caption{Consistency of the community structure for error and correct responses as measured by JS distance.}
    \label{fig:consistency}
\end{figure}

\begin{figure*}[t]
    \centering
    \centerline{a) \hspace{.45\textwidth} b) \vspace{0.5em}}
    \begin{minipage}[t]{\textwidth}
        \includegraphics{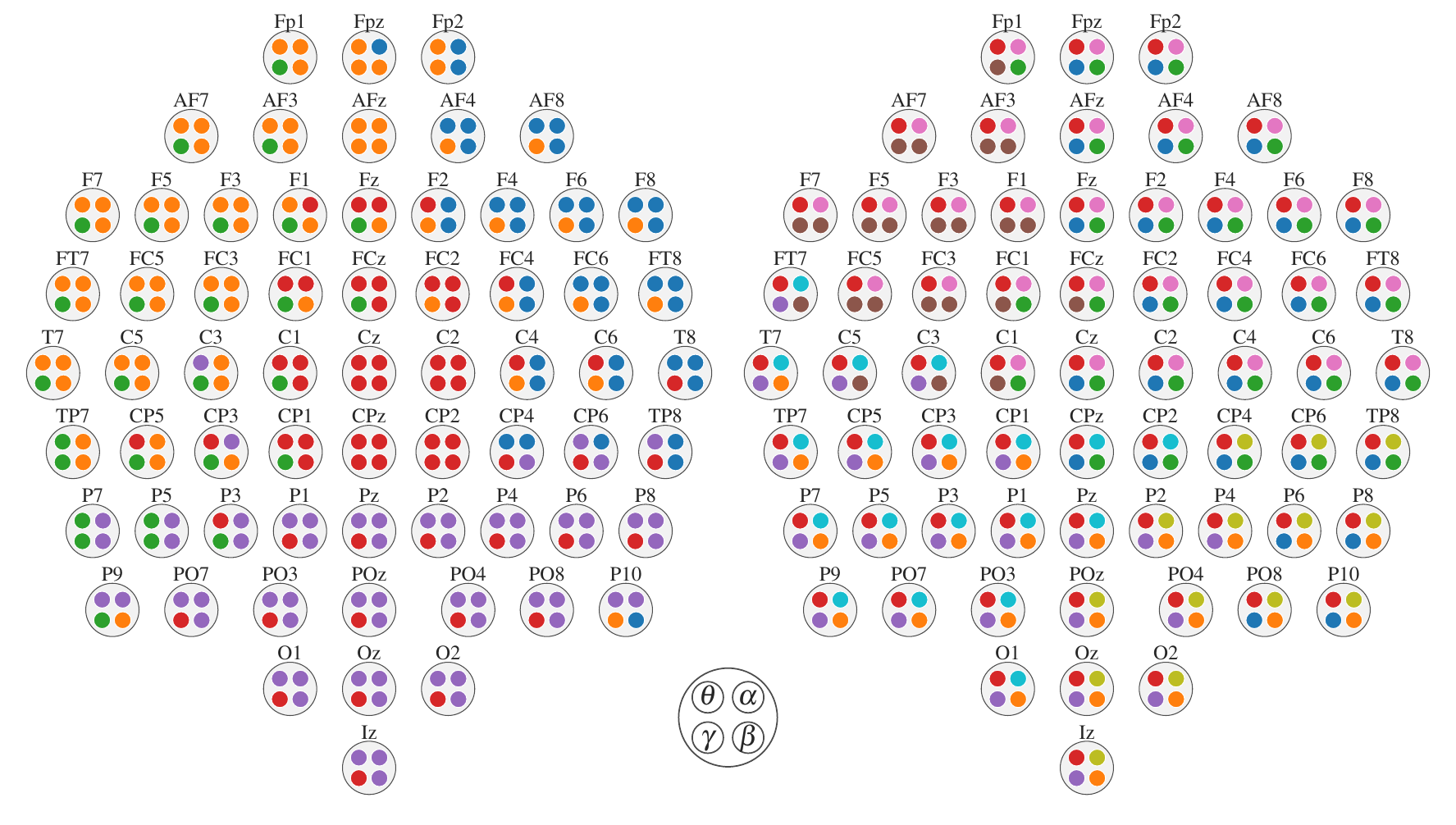}
    \end{minipage}
    \begin{minipage}[t]{\textwidth}
        \vspace{-2em}
        \centerline{c)}
    \end{minipage}
    \caption{Multilayer community structures for error (a)) and correct (b)) responses. Each electrode is shown with a circle that includes 4 colored circles, representing communities of electrodes across 4 frequency bands. c) shows the correspondence of inner circles to frequency bands. }
    \label{fig:group_comm}
\end{figure*}

After obtaining the community structure for each subject and both response types at the respective optimal $\gamma$ and $\omega$ values, we assess the consistency of community structures across subjects within each response type. We employed the multiview graph constructed in Section \ref{ssec:group_community_structure}, where we quantified the consistency of community structures across subjects by computing the similarity between the layers. In particular, we used Jensen-Shannon (JS) distance, which quantifies the similarity of layers based on their spectral properties \cite{de2016spectral}. To measure the similarity of layers $\layerh$ and $\layerk$ in the multiview graph, where each layer represents the co-clustering matrix of a subject, we  let $\mL^\layerh$ and $\mL^\layerk$ be the respective combinatorial Laplacian matrices of layers $\layerh$ and $\layerk$ and define $\widehat{\mL}^\layerh = \mL^\layerh/\trace(\mL^\layerh)$. JS distance is the square-root of JS divergence, which is defined as \cite{de2015structural}:
\begin{align}
    d_{JS}(\widehat{\mL}^\layerh, \widehat{\mL}^\layerk) = \frac{1}{2} d_{KL}(\widehat{\mL}^\layerh || \mM) + \frac{1}{2} d_{KL}(\widehat{\mL}^\layerk || \mM),
\end{align}
where $d_{KL}(\widehat{\mL}^\layerh || \mM) = \trace(\widehat{\mL}^\layerh(\log_2(\widehat{\mL}^\layerh) - \log_2(\mM)))$ is the Kullback-Liebler divergence and $\mM = (\widehat{\mL}^\layerh + \widehat{\mL}^\layerk)/2$. JS distance takes values between 0 and 1. We measured JS distance between each pair of subjects within each response type and plotted the average distance of each subject to other subjects in Fig. \ref{fig:consistency}. This plot shows that the average divergence for each subject with respect to the other subjects is lower for error response compared to the correct response. This indicates that there is more group level consistency in terms of topological organization for the error response. This is in line with prior work \cite{ozdemir2015hierarchical} that shows that the organization of the functional connectivity networks for correct response is similar to pre-stimulus networks. Thus, there is more variation across subjects for the correct response compared to response-evoked networks following an error response.

\subsection{Group Community Structure for Error and Correct Responses}

Once the optimal community structures are obtained for each subject and for each response type, the group community structure is extracted by using SC-ML described above. The number of communities is determined as the average of the number of communities detected for each subject. These values are 5 and 9 for the error and correct responses, respectively. Fig. \ref{fig:group_comm} illustrates the group community structure for error and correct responses for the multi-frequency networks. For error response, it can be seen that there is consistency between the community structure of the low frequency bands, e.g., $\theta$ and $\alpha$. We also note that during ERN there is a community comprised of the frontal-central nodes in the $\theta$-band. This community structure is consistent with prior work that indicates the role of medial prefrontal cortex (mPFC) during ERN in the $\theta$-band \cite{ozdemir2015hierarchical}. However, the proposed approach shows that this is an across-frequency community as the frontal-central regions from the low-frequency bands are in the same community as the parietal-occipital regions from the $\gamma$-band. While this is a new finding in terms of network organization, our recent work indicates that the phase of the $\theta$ band oscillations from the frontal-central regions modulate the amplitude of the $\gamma$ band oscillations in the parietal-occipital regions following an error response \cite{munia2021multivariate}. Prior studies also hypothesized that error-related negativity initiates the medial frontal based top-down control mechanisms to improve the performance after an error
response \cite{holroyd2002neural}. Thus, the communities detected are consistent with previous
literature reflecting higher theta-gamma coupling in the medial
frontal cortex and relating this with error-related negativity.

The community structure for the correct response is mostly within-layer indicating the lack of coupling across different frequency bands. Our prior work comparing PAC between response types supports this observation as there is significantly higher cross-frequency coupling during error monitoring \cite{munia2019time}.

The similarity of the group structures for error and correct responses was quantified using three different metrics: Normalized Mutual Information (NMI), Adjusted Rand Index (ARI) and Variation of Information (VI). NMI and ARI are normalized and a value close to 1 indicates similarity, while for VI small values indicate similarity. The similarity is computed for each frequency band. From Table \ref{tab:similarity}, it can be seen that the community structure in the $\gamma$ band has the highest similarity between the two response types while the community structures in the $\theta$ band are not similar at all. This is due to the fact that  during error monitoring, there is increased phase synchronization in the $\theta$ band. Thus, the community structure following an error response is different from the correct response in this frequency band. 

\begin{table}[h]
    \centering
    \caption{Similarity of community structures between Error and Correct Responses Across Frequency Bands}
    \begin{tabularx}{\columnwidth}{cYYYY}
        \toprule
        & $\theta-\theta$ & $\alpha-\alpha$ & $\beta-\beta$ & $\gamma-\gamma$ \\
        \cmidrule(lr){2-5}
        NMI & $0.0$ & $0.36$ & $0.42$ & $0.44$ \\
        ARI & $0.0$ & $0.22$ & $0.34$ & $0.29$ \\
        VI & $1.47$ & $1.53$ & $1.38$ & $1.18$ \\
        \bottomrule
    \end{tabularx}
    \label{tab:similarity}
\end{table}

\section{Conclusions}

This paper introduced a multilayer model of functional connectivity of the brain. In particular, we provided a data-driven way to construct multi-frequency connectivity networks where layers correspond to different frequency bands. The resulting networks capture both within frequency connectivity and cross-frequency coupling in a single framework. We then introduced a new definition of modularity for multilayer networks such that the null model preserves the heterogeneity of edge weights across layers. The community detection algorithm resulting from the maximization of this new multilayer modularity function is applied to EEG data collected during error monitoring. The results indicate that following an error response, the brain organizes itself to form communities across frequencies, in particular between theta and gamma bands. This cross-frequency community formation is not observed for the correct response which indicates that the cross-frequency coupling is primarily associated with cognitive control. Moreover, we observed that the community structures detected for the error response were more consistent across subjects compared to the community structures for correct response.

Future work will consider extension of this multilayer model to higher dimensions, e.g. multi-aspect multilayer brain networks such as temporal multi-frequency connectivity networks. Compared to current work where subjects' community structure is found separately and then combined through multiview community detection, future work can use multi-aspect multilayer networks constructed from subjects' multilayer networks. This approach will allow simultaneous detection of communities of subjects similar to \cite{betzel2019community}. Future work will also consider different null models in the definition of modularity such as the constant Potts model, which is shown to be resolution limit free \cite{traag2011narrow}. Finally, in this work we aimed to find the optimal resolution and inter-layer scale parameter; future work can focus on a multi-scale approach where the aim is to combine community structures from different resolutions and inter-layer scales \cite{jeub2018multiresolution}.

\section{Acknowledgement}

The authors would like to thank Dr. J. S. Moser from Michigan State University for providing the EEG data.

\bibliographystyle{ieeetr}
\bibliography{refs}

\end{document}